\documentclass[pdflatex,sn-mathphys-num]{sn-jnl}% Math and Physical Sciences Numbered Reference Style
\usepackage{graphicx}%
\usepackage{multirow}%
\usepackage{amsmath,amssymb,amsfonts}%
\usepackage{amsthm}%
\usepackage{mathrsfs}%
\usepackage[title]{appendix}%
\usepackage{xcolor}%
\usepackage{textcomp}%
\usepackage{manyfoot}%
\usepackage{booktabs}%
\usepackage{algorithm}%
\usepackage{algorithmicx}%
\usepackage{algpseudocode}%
\usepackage{listings}%
\newcommand{\pfr}[2]{\ensuremath{\frac{\partial #1}{\partial #2}}}
 
\newcommand{\mb}[1]{\mathbf{#1}}
\newcommand{\gb}[1]{\boldsymbol{#1}}

\newcommand{\tr}{\mathrm{tr}\,}
\newcommand{\Qvec}{\mathbf{Q}}

\begin{document}

\title[Isosceles nematic wells]{Nematic equilibria in isosceles triangles: The effects of edge length and apex angle on solution landscapes in a reduced Landau--de Gennes framework}

\author{\fnm{Prabakaran} \sur{Rajamanickam}$^1$}

\author{\fnm{Yucen} \sur{Han}$^2$}

\author{\fnm{Thuriya} \sur{Alhinai}$^3$}

\author{\fnm{Apala} \sur{Majumdar}$^1$}

\affil{$^1$\orgdiv{Department of Mathematics}, \orgname{University of Manchester}, \orgaddress{\city{Manchester}, \postcode{M13 9PL}, \country{United Kingdom}}}

\affil{$^2$\orgdiv{Center for Applied Mathematics}, \orgname{Renmin University of China}, \orgaddress{\city{Beijing}, \postcode{100872}, \country{China}}}

\affil{$^3$\orgdiv{Department of Mathematics and Statistics}, \orgname{University of Strathclyde}, \orgaddress{\city{Glasgow}, \postcode{G1 1XQ}, \country{United Kingdom}}}

\abstract{We study equilibrium configurations of nematic liquid crystals confined to two-dimensional isosceles triangles, subject to tangent boundary conditions. This toy problem is motivated by the effects of geometrical asymmetry on equilibria in variational problems arising in liquid crystal theory. There are two key geometrical parameters for an isosceles triangle - the triangle edge length and the apex angle. The nematic equilibria are modelled by minimizers of a reduced Landau-de Gennes free energy in this setting. For small edge lengths, we provide a universal, angle-based local classification of nematic equilibria near the vertices as to whether the nematic director exhibits a splay, bend or singular profile depending on the vertex opening angle. In the large domain limit, we demonstrate the existence of multiple competing nematic equilibria -- the three rotated solutions, for which the nematic director bends between a pair of adjacent vertices, and a \emph{trefoil} solution featuring an interior point defect. For acute apex angles, we show that the trefoil solution is stable for small edge lengths.  The interior point defect of the trefoil solution migrates to one of the base vertices, as the edge length increases, and is finally expelled giving way to the rotated solutions, if the apex angle is small enough. Our numerical results suggest that there is a unique trefoil solution on the equilateral triangle for all edge lengths, and a unique rotated solution on isosceles triangles with wide apex angles. These results  yield interesting insight into how geometrical asymmetry can tailor equilibria and self-assembly processes in confined nematic systems.}


%%\pacs[JEL Classification]{D8, H51}

%%\pacs[MSC Classification]{35A01, 65L10, 65L12, 65L20, 65L70}

\maketitle

\section{Introduction}\label{sec1}

Nematic Liquid Crystals (NLCs) are mesogenic materials that combine fluidity with some ordering characteristics of crystalline solids, i.e., the constituent rod-like nematic molecules have no positional order but exhibit long-range orientational order~\cite{de1993physics}. The orientational order manifests in distinguished directions of average molecular alignment, referred to as \emph{nematic directors}. Consequently, NLCs have directional or anisotropic optical, mechanical and rheological responses to external stimuli, such as external fields, mechanical stress and incident light, with the \emph{directors} acting as special material directions. This intrinsic anisotropy naturally lends itself to multi-faceted applications of NLCs across electro-optic devices, photonics, robotics, smart materials, healthcare technologies, and many more~\cite{lagerwall2012new}. NLCs in confinement are a dynamic playground to study the interplay between geometry, anchoring, material properties and temperature in pattern formation, namely, the multiplicity, structural details and singular sets of observable nematic equilibria and how these equilibria can be manipulated for tailor-made applications. \emph{Multistable NLC systems} are a topic of contemporary interest, i.e., confined NLC systems with multiple observable equilibria that can act as multiple modes of functionality. In other words, multistable NLC systems are great candidates for smart systems aimed at understanding the fundamental mechanisms of pattern formation and self-assembly in NLCs. Typical examples of multistable NLC systems include display-type devices such as the planar bistable nematic device reported in~\cite{tsakonas2007multistable}, the Zenithally Bistable Nematic device~\cite{spencer2010zenithal}, droplets with multiple stable defect structures~\cite{noh2020dynamic} and liquid crystalline biological systems that exhibit multiple stable conformations of viruses and bacteria~\cite{tarafder2025targeted}. 

Real-life NLC systems are typically three-dimensional (3D) but quasi-2D (two-dimensional) NLC systems are widely used as building blocks for tessellated systems, colloidal systems, photonics, layered or lamellar systems~\cite{oladepo2022development, smalyukh2018liquid, lin2022formation}, etc. From a theoretical perspective, we need to understand NLC pattern formation in two-dimensions before progressing to three-dimensional studies of confined NLC systems. For example, 2D equilibria can act as building blocks for 3D equilibria in confined NLC systems~\cite{han2023reduced}. The 2D equilibria can also contain cues about transition pathways between 3D equilibria. In other words, 2D studies can provide vital information about the energy landscapes of real 3D systems. From an applications perspective, advances in materials technology mean that thin, flexible and portable (almost) 2D NLC systems are a reality, and hence it is important to classify nematic equilibria on canonical 2D systems.

In this paper, we mathematically model nematic equilibria within shallow 3D wells with an isosceles triangle cross-section, in the thin film limit or the limit of vanishing well thickness, within the celebrated continuum Landau--de Gennes (LdG) framework. The LdG theory is a powerful and widely used continuum theory for modelling nematic phases; it describes the nematic phase by a macroscopic order parameter--the LdG $\mathbf{Q}_f$-tensor order parameter which can be interpreted in terms of experimental measurables such as the dielectric anisotropy and birefringence data~\cite{patranabish2021quantum}. Mathematically speaking, the LdG $\mathbf{Q}_f$-order parameter is a symmetric, traceless $3\times 3$ matrix whose eigenvectors model the nematic directors, and the corresponding eigenvalues measure the degree of orientational order about these directors, respectively~\cite{majumdar2010landau}. The LdG theory is a variational theory with an associated free energy, typically of the form
\[
F_{3D}[\Qvec_f]: = \int_{\Sigma} f_{el}\left(\Qvec_f, \nabla \Qvec_f \right) + f_b\left(\mathbf{Q}_f\right)~dV,
\]
where $\Sigma$ is the 3D well, $f_b$ is a non-convex bulk potential that dictates temperature-driven phase transitions in homogeneous systems and $f_{el}$ is an elastic energy density that penalises spatial inhomogeneities. The elastic energy density is usually a quadratic and convex function of $\nabla \Qvec_f$, so that the energy density is lower semi-continuous and coercive. The physically observable NLC equilibria are modelled by global or local minimizers of the LdG free energy, subject to imposed boundary conditions and constraints~\cite{majumdar2010landau}. 

We work with planar boundary conditions on the top and bottom well surfaces, so that the nematic molecules are constrained to be in the plane of the cross-section, complemented by Dirichlet planar boundary conditions on the lateral surfaces. The Dirichlet conditions on the lateral surfaces are tangent to the edges of the isosceles triangle cross-section and invariant across the height of the well. In the limit of vanishing well thickness or height, the nematic equilibria are expected to be planar, i.e., independent of the $z$-coordinate or the vertical coordinate, and hence, invariant across the height of the well, so that they are fully determined by a reduced 2D problem defined on an isosceles triangular domain. This defines the reduced LdG (rLdG) framework for quasi-2D settings, as described in the next section, within which the planar nematic phase is modelled by a rLdG order parameter and the corresponding equilibria are modelled by minimizers of a suitably defined rLdG energy. The defect set is identified with the nodal set of the rLdG order parameter. 

In recent years, there has been substantial theoretical and numerical work on NLC equilibria in 2D settings. For example, in~\cite{DAVIDSON_MOTTRAM_2012}, the authors use conformal mapping techniques to analytically construct nematic director profiles in wedge-shaped 2D geometries, in the large-domain limit of the LdG theory. In~\cite{blow2013interfacial}, the authors study nematic configurations within rectangular grooves, with an emphasis on wetting and interfacial phenomena. NLC equilibria have also been extensively studied on polygonal domains. In~\cite{tsakonas2007multistable,luo2012multistability}, the authors study NLC equilibria on square domains with tangent boundary conditions in the rLdG framework and report the existence of two classes of stable equilibria on large square domains: the \emph{diagonal} solution, for which the planar nematic director aligns along one of the square diagonals and the \emph{rotated} solution, for which the planar nematic director rotates by $\pi$ radians between a pair of parallel square edges. The diagonal and rotated solutions exhibit long-term stability without external fields, and hence this is a prototypical bistable system. In~\cite{han2020reduced}, the authors carry out a systematic study of NLC equilibria on regular polygons, focussing on the effects of polygon symmetry and polygon size on the stable equilibria within the rLdG setting. Notably, there is a unique stable equilibrium configuration for sufficiently small polygons. This unique equilibrium has a fractional interior point defect on equilateral triangles, a pair of orthogonal line defects on square domains and an interior vortex defect for all remaining polygons. In contrast, there can be at least $K(K-1)/2$ stable equilibria on regular $K$-polygons with $K$ edges, for sufficiently large polygons. For example, there can be at least $15$ distinct NLC equilibria on large hexagons ($K=6$). The authors find that the polygon symmetry plays a crucial role in the singular set of the stable equilibria, and polygon size tunes multistability or the number of stable equilibria in the rLdG setting. In~\cite{han2021solution}, the authors shift focus to both energy minimizers and saddle points of the rLdG energy on regular hexagons with tangent boundary conditions. Saddle points can model unstable transient configurations, but they play a crucial role in non-equilibrium phenomena, such as switching between different stable equilibria and the selection of stable solutions in multistable systems. Saddle points of the rLdG energy are often characterised by intricate and symmetric defect constellations. It is indeed a mathematical puzzle as to how these defect arrangements control the modes of instability and the connectivity of the rLdG energy landscape; see also~\cite{han2019transition}.  

Polygonal domains are especially useful because they can be regarded as fundamental units for self-assembly processes. Regular polygons have substantial symmetry, and symmetry breaking can affect the associated solution landscapes. In this regard, there has been work on NLC equilibria on rectangular domains, and their connections to solution landscapes on square domains~\cite{yin2020construction, fang2020surface,shirectangle2022}. There have been studies of both energy minimizers and saddle points of the rLdG energy on rectangular domains, their dependence on the geometrical aspect ratio, and transition pathways mediated by the ejection of \emph{topological bubbles}. Among polygonal shapes, triangular domains hold a special place. They represent the simplest setting in which geometric frustration is unavoidable and the degree of frustration is typically stronger than in other regular polygons, which could offer fewer prospects for multistability. Despite their simplicity, NLC equilibria in triangular domains have not been studied as extensively as squares or hexagons (see, for example, the Table I in~\cite{yao2022defect}), even though triangles offer an ideal environment for studying the interplay between geometric frustration, defects and energetics.

In this paper, we study NLC equilibria on isosceles triangular domains in the rLdG setting, with Dirichlet tangent boundary conditions on the triangle edges. The tangent boundary conditions require the nematic director to be tangent to the edges. This means that the nematic molecules, on average, are constrained to be tangent to the triangle edges, inducing natural mismatches or singularities at the vertices. The isosceles triangle is parameterised by two geometrical parameters: a rescaled length $\lambda$ and an apex angle, denoted by $2\alpha$. We work at a fixed low temperature so that temperature-driven effects do not dictate pattern formation. The equilateral triangle is a special case with $2\alpha = 60^\circ$. We study two asymptotic limits - the $\lambda \to 0$ limit and the $\lambda \to \infty$ limit. The rLdG energy has a unique critical point, and hence unique minimizer in the $\lambda \to 0$ limit. Unlike the work in~\cite{han2020reduced}, we do not compute the limiting solution in the $\lambda \to 0$ limit. Instead, we provide a local description of the rLdG order parameter near each vertex, according to the vertex angle, in the $\lambda \to 0$ limit. This is a versatile description that can be employed for any geometry with sharp corners, based on similarity solutions. The local vertex profiles suggest that the limiting solution has an interior point defect for $2\alpha < 90^\circ$, and  an (interior) defect-free bend profile around the apex for $90^\circ < 2\alpha < 180^\circ$. More details are given in Section~\ref{sec:smalldomain}. This demonstrates how the unique equilibrium depends on the apex angle in the $\lambda \to 0$ limit. The numerical results show that the interior defect is a fractional $-\tfrac{1}{2}$ defect that migrates from near the base to the apex, as $2\alpha$ increases from $0^\circ$ to $90^\circ$, in the $\lambda \to 0$ limit. Thus, the apex angle also affects the defect location in this regime. In the $\lambda \to \infty$ limit (see Section~\ref{sec:largedomain}), we use asymptotic arguments to identify a set of four candidates for rLdG energy minimizers: three rotated solutions, for which the planar nematic director bends between a pair of adjacent vertices and a \emph{trefoil} solution featuring an interior fractional point defect.  We expect to have three energetically degenerate rotated solutions for equilateral triangles, and two energetically degenerate rotated solutions for $2\alpha \neq 60^\circ$, which are reflections of each other about the apex bisector. For small values of $2\alpha$, the trefoil solution is numerically observed to be energy minimizing for small $\lambda$, followed by a transition to one of the energetically degenerate rotated solutions for large $\lambda$. We numerically find the trefoil solution to be energy minimizing on equilateral triangles, for all $\lambda$. This is an interesting observation of an energy minimizer with a stable interior fractional defect that can act as a binding site for self-assembly processes. For large values of $2\alpha$, say $2\alpha \geq 90^\circ$, the energy minimizer is the unique rotated solution that bends around the wide apex angle. Hence, isosceles triangles, with small values of $2\alpha$, can offer richer solution landscapes compared to equilateral triangles or isosceles triangles with wide apex angles, as suggested by the numerical results in Section~\ref{sec:numerical}. Our study, while simple, provides useful insights into how NLC equilibria in triangular domains can be used as building blocks for NLC equilibria in more complex polygons. For example, the diagonal solution in square domains  can be regarded as a composition of two rotated solutions in two right-angled isosceles triangles. In fact, Han \textit{et al.}~\cite{han2021solution} have recently identified a new  critical point of the rLdG energy on a hexagon, referred to as the T-state, which may be interpreted as a composite solution configured from the rotated and trefoil solutions described in this paper. We speculate that the rotated and trefoil solutions can also be used to construct critical points of the rLdG energy on parallelograms with tangent boundary conditions. We conclude with some remarks on the implementation of the tangent boundary conditions and analytic recipes for families of critical points of the rLdG energy in Section~\ref{sec:conclusions}.

\section{Problem description} \label{sec:problem}

\begin{figure}[h!]
\centering
\includegraphics[width=0.4\textwidth]{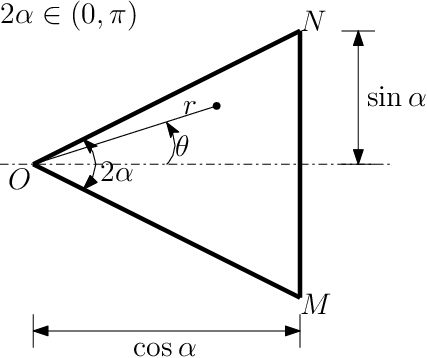} 
\caption{Geometry of the LC sample: Isosceles triangle $\Delta OMN$ of leg length unity and apex angle $2\alpha\in(0,\pi)$.} 
\label{fig:triangle}
\end{figure}

Consider the three-dimensional well domain: $\Sigma = \Omega \times (0, H)$, where $\Omega$ is the two-dimensional (2D) cross-section and $H$ is the well height. We take $\Omega$ to be an isosceles triangle with leg length $h$ and apex angle $2\alpha \in (0, \pi)$; its base has length $2h\sin\alpha$ and its height is $h\cos\alpha$. After rescaling the coordinates by $h$, the domain $\Omega:=\Delta OMN$ is represented as shown in Fig.~\ref{fig:triangle}. We impose tangent or planar degenerate boundary conditions on $z=0, H$, so that the nematic molecules are constrained to be in the plane of these surfaces. In what follows, we assume that $H$ is much smaller than the cross-sectional dimensions of $\Omega$.

We work in the Landau-de Gennes (LdG) framework for modelling nematic phases in confinement. In the conventional LdG theory, the nematic state is modelled by the LdG tensor order parameter, which is a symmetric and traceless $3\times 3$ matrix, with five degrees of freedom. Following Theorem 5.1 in~\cite{golovaty2015dimension} (also see Theorem 2.1 in~\cite{wang2019order}), we deduce that the physically relevant LdG order parameters (modelled by energy minimizers) have a fixed eigenvector, $\mb e_z$ (the unit-vector in the $z$-direction) with associated fixed eigenvalue, in the thin-film limit or in the $H\to 0$ limit and with planar degenerate or tangent boundary conditions on $z=0, H$. This necessarily reduces the number of degrees of freedom from five in the full framework, to two in the reduced LdG framework.

In other words, in the thin-film limit, the domain reduces to $\Omega$ and the macroscopic nematic phase can be modelled within a reduced two-dimensional Landau--de Gennes (rLdG) framework~\cite{han2020reduced}. In the rLdG framework, the nematic order parameter is the $z$-invariant rLdG $\mb Q$-tensor order parameter in the space
\[
S_2 = \{\mb Q \in \mathbb R^{2 \times 2}\mid \mb Q = \mb Q^T,  \tr\mb Q = 0\}, 
\] 
i.e., the rLdG $\mb Q$-tensor order parameter is a symmetric, traceless $2\times 2$ matrix with two degrees of freedom. Consequently, any admissible $\mb Q$ can be written as
\begin{equation}\label{eq:Q}
    \mb Q =  s(\mb n\otimes\mb n- \tfrac{1}{2}\mb I) = \begin{bmatrix}
        q_1 & q_2 \\ q_2 & -q_1
    \end{bmatrix} 
\end{equation}
where $\mb n=(\cos\varphi,\sin\varphi)^T$ is the nematic director or the eigenvector with the positive eigenvalue, $\varphi$ is the director angle measured from the $x$-axis and $s$ is the non-negative scalar order parameter describing the degree of molecular alignment along $\mb n$ (in an averaged sense). In other words, $\mb n$ models the preferred direction of averaged molecular alignment in $\Omega$ and the planar defect set is captured by the nodal set of $s$. Both $s$ and $\mb n$ only depend on the planar coordinates defined on $\Omega$ and are hence, invariant in the $z$-direction. These variables are related to $(q_1, q_2)$ by
\begin{eqnarray}
    && q_1(x,y) = \tfrac{1}{2}s(x,y)\cos2\varphi(x,y), \quad q_2(x,y) = \tfrac{1}{2}s(x,y)\sin2\varphi(x,y), \nonumber \\ && s(x,y)=2\left(\sqrt{q_1^2+q_2^2}\right)(x,y), \quad \varphi(x,y) = \frac{1}{2}\tan^{-1}\frac{q_2(x,y)}{q_1(x,y)}, \quad (x,y)\in \Omega.  \label{q1q2}
\end{eqnarray} 
The isotropic phase is modelled by $\mb Q = 0$, while orientationally ordered nematic phases are modelled by $\mb Q \neq 0$. We recall that the mapping between the full LdG $\mb Q$-tensor, denoted by $\mb Q_f$ below, and the rLdG $\mathbf{Q}$-tensor defined in~\eqref{eq:Q}, is given by
\begin{equation}
    \mathbf{Q}_f=
            \begin{pmatrix}
            q_1-q_3 & q_2 & 0\\
            q_2 & -q_1-q_3 & 0\\
            0 & 0 & 2q_3
        \end{pmatrix},
\end{equation}
where $q_1$ and $q_2$ are as above and $q_3$ is a fixed known constant determined by the temperature and boundary conditions (coded in terms of surface energies). There is no notion of biaxiality in the rLdG framework or in the 2D framework, since there can only be one distinguished planar director in a 2D setting.
  
The rLdG free energy per unit film thickness is given by
\begin{equation}
    F[\mb Q] = \int_\Omega \left[\frac{L}{2}|\nabla\mb Q|^2 + \frac{A}{2}\tr\mb Q^2 + \frac{C}{4}(\tr\mb Q^2)^2\right]d^2x, \label{Feq}
\end{equation}
where $d^2x$ is the area element on $\Omega$, the parameter $A=\alpha (T - T^*)$ depends linearly on temperature and the material constants, $\alpha$, $L$ and $C$ are positive~\cite{han2020reduced}. 
In spatially homogeneous systems, $F$ is globally minimized by
\begin{equation}
    \mb Q_{hom} = \begin{cases}
        \mb 0  & \quad \text{for  } A>0,\\
        s_+ (\mb n_+\otimes\mb n_+-\tfrac{1}{2}\mb I)  & \quad \text{for  } A<0,
    \end{cases}     
\end{equation}
where $s_+=\sqrt{-2A/C}$ and $n_+\in \mathbb S^1$ is an arbitrary unit vector in the $xy$-plane. In what follows, we fix $A<0$ and work in the ordered nematic regime, focusing on minimizers of~\eqref{Feq} on $\Omega$ and their dependence on $h$ and $\alpha$. The energy minimizers mathematically model the physically observable equilibrium nematic configurations, for this benchmark problem.  

%In this article, we study the minimizers of the free energy in Eq.~\ref{Feq} in the nematic regime at a fixed temperature with $A < 0$. Our interest lies in geometrically frustrated domains for which equilibrium configurations are intrinsically inhomogeneous. The domain under consideration is an isosceles triangle with leg length $h$ and apex angle $2\alpha \in (0, \pi)$; its base has length $2h\sin\alpha$ and its height is $h\cos\alpha$. After rescaling the coordinates by $h$, the domain $\Omega:=\Delta OMN$ is represented as shown in Fig.~\ref{fig:triangle}. 

We rescale lengths by $h$, the order parameter $\mb Q$ by $s_+$, and the  energy by $s_+^2 L$. With these choices, the rescaled energy becomes
\begin{align}
     F[\mb Q] = \int_\Omega \left\{\frac{1}{2}|\nabla \mb Q|^2+ \frac{\lambda^2}{4}((\tr\mb Q^2)^2-\tr\mb Q^2) \right\} d^2x, \label{freeenergy}
\end{align}
where $\lambda=hs_+\sqrt{C/L}$ is the non-dimensional domain size, since $A, C$ and $L$ are kept fixed in this study. The corresponding Euler--Lagrange equations for $q_1$ and $q_2$ are:
\begin{align}
    \nabla^2 q_1 = \tfrac{1}{2}\lambda^2 q_1(4q_1^2+4q_2^2-1), \label{q1}\\  \nabla^2 q_2 = \tfrac{1}{2}\lambda^2 q_2(4q_1^2+4q_2^2-1).   \label{q2}
\end{align}
We impose strong tangential anchoring that are modelled by the following Dirichlet conditions for $(q_1, q_2)$ on $\partial \Omega$:
\begin{align}
r\in(0,1),\,\theta=+\alpha: & \qquad q_1 = +\tfrac{1}{2}\cos2\alpha, \quad q_2 = +\tfrac{1}{2}\sin2\alpha, \label{BC1}\\
   r\in(0,1),\,\theta=-\alpha: & \qquad q_1 = +\tfrac{1}{2}\cos2\alpha, \quad q_2 = -\tfrac{1}{2}\sin2\alpha, \label{BC2}\\   
   r=\frac{\cos\alpha}{\cos\theta},\,\theta\in(-\alpha,+\alpha): & \qquad q_1 = -\tfrac{1}{2}, \qquad \quad \,\,\,\,  q_2 =0. \label{BC3}
\end{align}
These boundary conditions naturally induce geometric frustration and spatially inhomogeneous nematic equilibrium configurations.
Following~\cite{han2020reduced}, the discontinuity of $\mb Q$ at the vertices is regularised by introducing a small neighbourhood of size $\epsilon\ll 1/2$ along each edge, where $\epsilon$ denotes the distance measured from the vertex along the edge. On this neighbourhood, the boundary data are smoothed using a linear interpolation.

\noindent The details of the numerical methods, as used in this paper, are given in Appendix A.

\subsection{Formulation of the stability analysis}

Let $\bar{\mb q}=(\bar q_1,\bar q_2)^T$ be a critical point of the rescaled rLdG free energy~\eqref{freeenergy}, that satisfies the Dirichlet boundary conditions in~\eqref{BC1}-\eqref{BC3}. We assess the stability of $\bar{\mb q}$ by means of the associated gradient-flow dynamics
\begin{equation}
    \pfr{\mb q}{t} = -\frac{\delta F}{\delta \mb q}, \quad \text{where} \quad \mb q= \begin{bmatrix}
        q_1 \\ q_2,
    \end{bmatrix} 
\end{equation}
and $t$ is a fictitious time that governs the evolution of the system along a path of decreasing energy. By superposing small perturbations on $\bar{\mb q}$ of the form,
\begin{equation}
   \mb q = \bar{\mb q}(x,y) + \hat{\mb q}(x,y) e^{-\sigma t},
\end{equation}
with $|\hat{\mb q}|\ll | \bar{\mb q}|$, and linearising the gradient-flow equations about $\bar{\mb q}$, we obtain the eigenvalue problem~\cite{rajamanickam2026strong}
\begin{equation}
    \mb H_{\bar{\mb q}}  \hat{\mb q}  = \sigma \hat{\mb q},  \qquad 
    \mb H_{\bar{\mb q}} =  -2\mb I \nabla^2 +\lambda^2[8\bar{\mb q}\otimes\bar{\mb q} + (4|\bar{\mb q}|^2-1)\mb I ],
 \label{eigenvalue}
\end{equation}
complemented by homogeneous Dirichlet conditions  $\hat{\mb q} =\mb 0$ on $OMN$. Here, $\mb H_{\bar{\mb q}}$ is the Hessian operator of the rLdG energy~\eqref{freeenergy} evaluated at the critical point and is defined via the second variation
\begin{equation}
\delta^2 F[\bar{\mb q}](\gb\eta,\gb\psi)
= \int_\Omega \gb\eta(\mb x)^T  \mb H_{\bar{\mb q}} \gb\psi(\mb x)  d^2x,
\end{equation}
where $\gb\eta,\gb\psi \in H_0^1(\Omega;\mathbb R^2)$ are admissible test functions. Equivalently, one could formally write the double-integral form
\begin{align}
    \delta^2 F[\bar{\mb q}](\gb\eta,\gb\psi) 
= \iint_\Omega 
\gb\eta(\mb x)^T 
\left. \frac{\delta^2 F}{\delta \mb q(\mb x)\delta \mb q(\mb y)} \right|_{\bar{\mb q}} 
 \gb\psi(\mb y)  d^2y  d^2x
\\ \text{with} \qquad \frac{\delta^2 F}{\delta \mb q(\mb x)\delta \mb q(\mb y)} 
= \mb H_{\bar{\mb q}}  \delta(\mb x-\mb y),
\end{align}
since $F$ is a local functional and the delta function collapses the integral to the single integral above. Since $\mb H_{\bar{\mb q}}$ is self-adjoint, all its eigenvalues are real. The spectrum is discrete and may be arranged in non-decreasing order, $\sigma_1 \leq \sigma_2 \leq \sigma_3 \leq \dots$. The critical point $(\bar q_1,\bar q_2)$ is asymptotically stable when $\sigma_1>0$, marginally stable when $\sigma_1=0$ and unstable when $\sigma_1<0$. The Morse index is the number of negative eigenvalues and quantifies the number of linearly unstable directions in the solution landscape. The (global or local) energy minimizers are stable critical points with index-$0$. In particular, index-$1$ saddle points of~\eqref{freeenergy} are often referred to as \emph{transition states} and play a crucial role in transition pathways between distinct stable (index-$0$) critical points of~\eqref{freeenergy}~\cite{han2021solution}.

\section{The small-domain limit: the $\lambda\to 0$ limit}\label{sec:smalldomain}

\begin{figure}[h!]
\centering
\includegraphics[width=0.9\textwidth]{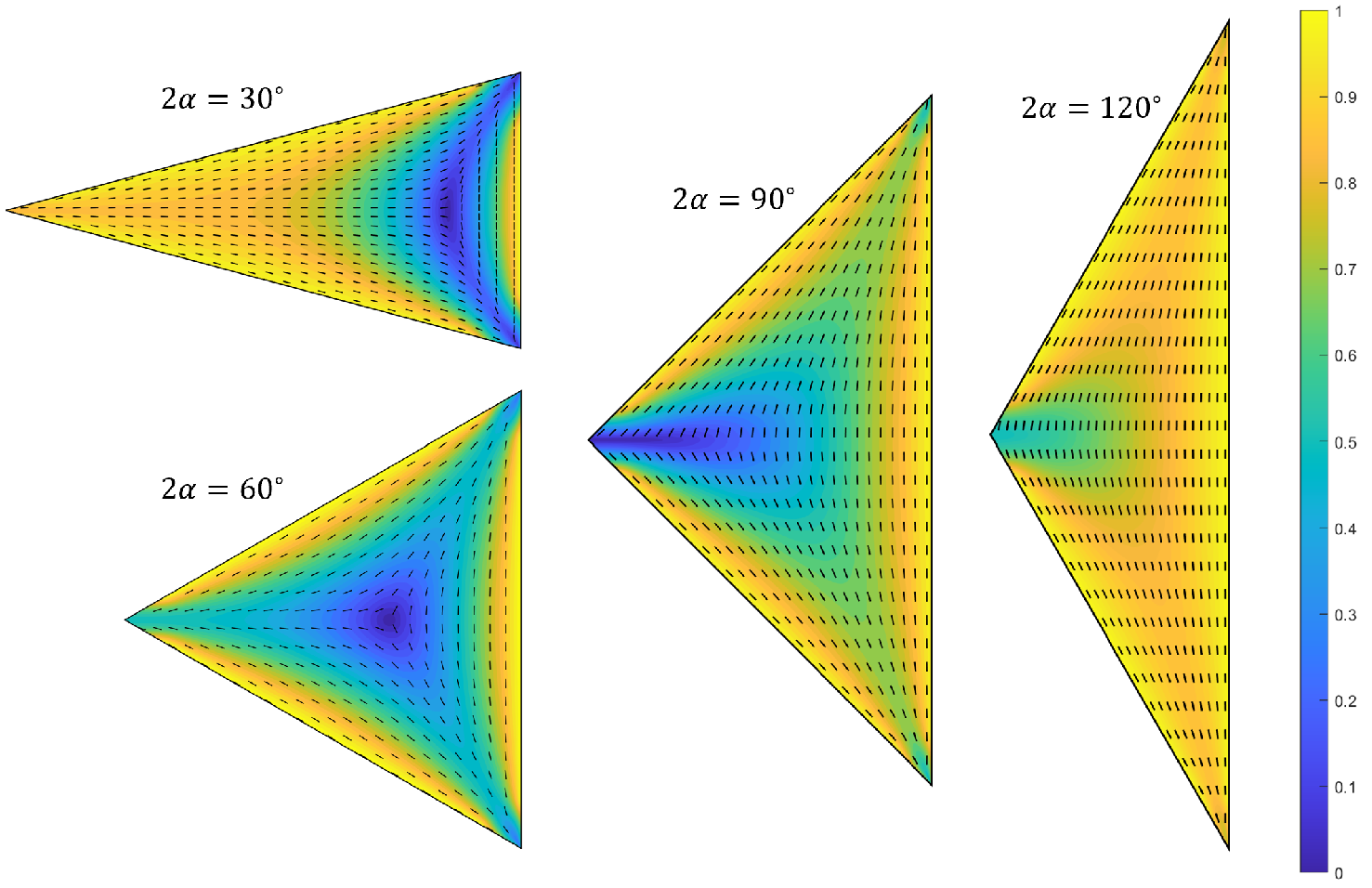}  
\caption{Representative solutions of \eqref{eq:lambda0} with tangent boundary conditions: the black lines label the nematic director field with $2\alpha = 30^\circ$, $2\alpha = 60^\circ$, $2\alpha = 90^\circ$ and $2\alpha = 120^\circ$. The colour bar corresponds to the scalar order parameter $s=2\sqrt{q_1^2+q_2^2}$.} 
\label{fig:directorlam0}
\end{figure}

There is a unique critical point (and hence, unique global minimizer) of~\eqref{freeenergy} in the $\lambda \to 0$ limit~\cite{han2020reduced}. In this limit, the Euler--Lagrange equations~\eqref{q1}-\eqref{q2} reduce to Laplace equations for $q_1$ and $q_2$,
\begin{equation} \label{eq:lambda0}
    \nabla^2 q_1 =0, \qquad \nabla^2 q_2 =0
\end{equation}
subject to the Dirichlet boundary conditions~\eqref{BC1}-\eqref{BC3} on $OMN$. One can use conformal mapping methods to construct analytic solutions of the Laplace equation on polygonal domains, as demonstrated in~\cite{han2020reduced}, and these ideas can also be adapted to the present problem. One may also attempt separation of variables, although the resulting eigenfunctions are usually non-orthogonal in isosceles triangles~\cite{sparrow1962laminar,damle2010understanding}. The equilateral and right-isosceles cases are the main exceptions, for which orthogonality is preserved.

We plot representative numerical solutions in Fig.~\ref{fig:directorlam0}. Recall that there is only one geometrical parameter, the apex angle denoted by $2\alpha$ in the rescaled problem, since the length $h$ is absorbed into $\lambda$. A vertex is labelled as a \emph{splay} vertex if $\mb n$ has a splay or radial profile near the vertex, to match the tangent boundary conditions on the intersecting edges. A vertex is labelled as a \emph{bend} vertex if $\mb n$ bends near a vertex, to match the tangent boundary conditions on the two intersecting edges. When $2\alpha=30^\circ$ and $2\alpha=60^\circ$, the nematic director profile has a  \textit{trefoil} or \textit{tribrachial} pattern with three splay corners and an interior $-\tfrac{1}{2}$ defect. In the right isosceles triangle, the base vertices are splay corners while the apex develops a discontinuity in the director field. When $2\alpha = 120^\circ$, the base vertices are splay corners whilst the apex vertex is a bend corner. These results agree with the numerical results reported in Fig.~7 of~\cite{han2020reduced}.

\begin{figure}[h!]
\centering
\includegraphics[width=0.5\textwidth]{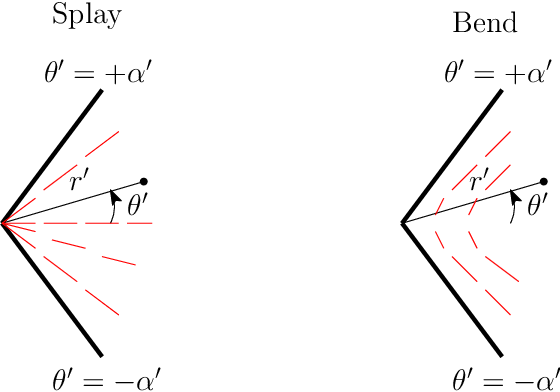} 
\caption{An arbitrary splay and bend corner with a corner angle $2\alpha'\in(0,2\pi)$, described in terms of the local polar coordinates $(r',\theta')$.} 
\label{fig:corner}
\end{figure}

To rationalize these numerical observations, consider a generic corner formed by the lines $\theta'=\pm\alpha'$; see Fig.~\ref{fig:corner}. The corner neighbourhood can be defined in terms of local polar coordinates, $(r',\theta')$, where $0< r'  \ll 1$ and $\theta' \in [-\alpha', \alpha' ]$.  
%$0< \epsilon<r'  \ll 1$
We can construct a similarity solution of \eqref{eq:lambda0} with tangent boundary conditions, for $r'\ll 1$, given by
\begin{align} \label{eq:similarity}
    q_1 & = \frac{1}{2}\cos2\alpha' + \sum_{n=0}^\infty A_n r'^{\mu_n}\cos\mu_n\theta', \qquad \mu_n=\frac{\pi}{\alpha'}(n+\tfrac{1}{2}), \\
    q_2 & = \frac{1}{2}\sin2\alpha'\, \frac{\theta'}{\alpha'} + \sum_{n=0}^\infty B_n r'^{\nu_n}\sin\nu_n\theta', \qquad \nu_n=\frac{\pi}{\alpha'}(n+1).
\end{align}
The expansion coefficients, $A_n$ and $B_n$, contain information about the global nature of the solution and cannot be determined from a purely local consideration. However, the leading-order terms for $r'\ll 1$ are completely prescribed, from which we deduce that
\begin{align}
s = \sqrt{\cos^22\alpha' + \frac{\theta'^2}{\alpha'^2}\sin^22\alpha'},\qquad
    \varphi=\frac{1}{2}\mathrm{atan2} \left(\frac{\theta'}{\alpha'}\sin2\alpha',\cos2\alpha'\right),
\end{align}
upon neglecting higher-order terms of order $r'^{\pi/2\alpha'}$. The second expression shows that $\varphi$ is undefined along the corner bisector $\theta' = 0$ precisely when $2\alpha' = \tfrac{\pi}{2}$ or $2\alpha' = \tfrac{3\pi}{2}$, and this corresponds to a defect in the director profile. In these two cases, 
\begin{align}
   \varphi = +\frac{\pi}{4}\mathrm{sgn}(\theta') \quad \text{as} \quad 2\alpha' \to \tfrac{1}{2}\pi  \qquad  \text{and} \qquad \varphi = -\frac{\pi}{4}\mathrm{sgn}(\theta') \quad\text{as} \quad
    2\alpha' \to \tfrac{3}{2}\pi.
\end{align}

\begin{table}[htbp]
\footnotesize
\caption{Local description of the nematic director profile near an arbitrary corner with a corner angle $2\alpha'$, with tangent boundary conditions on intersecting edges at the corner.} \label{tab:corner}
  \begin{tabular}{|c|c|c|} \hline
  Type & Corner angle & Solution character  \\ \hline
   I & $0^\circ<2\alpha'<90^\circ$ & Splay corner (splaying towards the bulk LC) \\
   II & $2\alpha'=90^\circ$ & Defect along the bisector (splay-to-bend transition) \\
  III&   $90^\circ<2\alpha'<180^\circ$ & Bend corner (bending around the corner) \\ 
  IV &  $2\alpha'=180^\circ$ & No distortion (reversal of bend direction) \\
  V &  $180^\circ< 2\alpha' < 270^\circ$ & Bend corner (bending towards the bulk LC) \\
  VI &  $2\alpha'=270^\circ$ & Defect along the bisector (bend-to-splay transition) \\
   VII & $270^\circ<2\alpha'<360^\circ$ & Splay corner (splaying towards the corner) \\ \hline
  \end{tabular}
\end{table}

\begin{figure}[h!]
\centering
\includegraphics[width=1\textwidth]{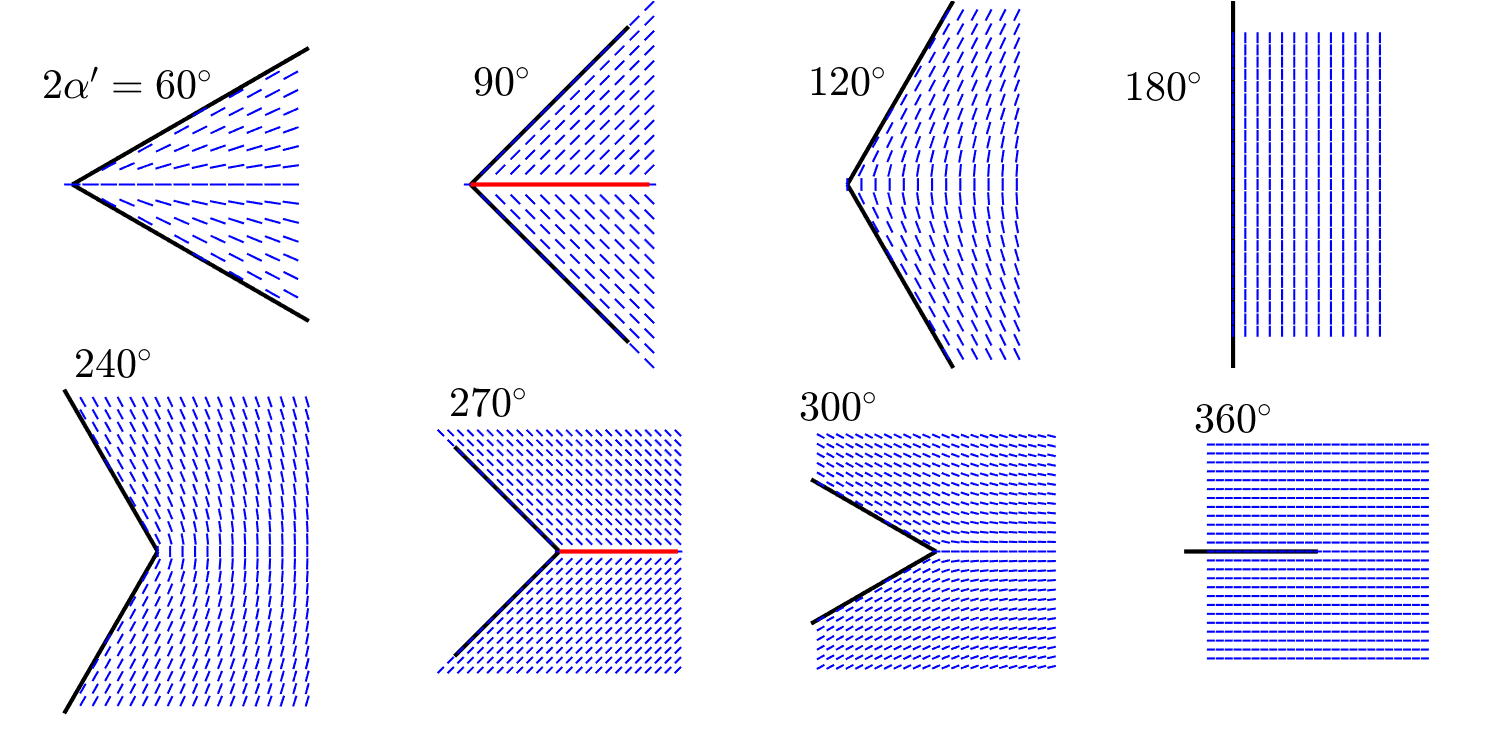} 
\caption{Plots of the similarity solution~\eqref{eq:similarity} for different choices of $2 \alpha'$. Defects emerge along $\theta'=0$ when $2\alpha'=90^\circ$ and $2\alpha'=270^\circ$.}  
\label{fig:cornerzero}
\end{figure}

The predictions of the similarity solution are summarized in Table~\ref{tab:corner} and Fig.~\ref{fig:cornerzero}, yielding a natural classification of vertices according to the vertex angle in the $\lambda\to 0$ limit. These predictions agree with the results in~\cite{han2020reduced} for regular polygons. In~\cite{han2020reduced}, the authors analytically construct solutions of the limiting equations~\eqref{eq:lambda0} on regular $K$-polygons with $K$ sides, with tangent Dirichlet conditions for $q_1$ and $q_2$, whereas the vertex classification in Table~\ref{tab:corner} provides a local director description near a vertex. The authors show that the unique solution of \eqref{eq:lambda0} has three splay corners and an interior $-\tfrac{1}{2}$-defect for $K=3$ (equilateral triangle); has two defect lines along the square diagonals for $K=4$ (square domain); and the unique solution of~\eqref{eq:lambda0} with tangent boundary conditions is the \emph{Ring} solution for $K>4$. The Ring solution has $K$ bend vertices with an interior central $+1$-point defect. The limiting solutions  constructed in~\cite{han2020reduced} are consistent with the predictions in Table~\ref{tab:corner}. According to Table~\ref{tab:corner}, the equilateral triangle has three type-I vertices with $2\alpha = 60^\circ$, i.e., the corresponding similarity solution has three splay vertices, in agreement with the limiting solution constructed in~\cite{han2020reduced}. A square has $4$ type-II vertices characterised by $2\alpha=90^\circ$, and there are defects along the vertex bisectors. The limiting solution on square domains (with tangent boundary conditions) is the Well Order Reconstruction Solution (WORS) of~\eqref{eq:lambda0}. The WORS is the unique solution of~\eqref{q1}-\eqref{q2}, for small $\lambda$, with two orthogonal defect lines (characterised by $s=0$) along the square diagonals~\cite{han2020reduced,fang2020surface,shi2022nematic}, consistent with the vertex classification in Table~\ref{tab:corner}.  The remaining regular $K$-polygons  have type-III vertices (pentagon, hexagon and so on), for which the similarity solution has a bend profile in agreement with the limiting \emph{Ring} solution reported in~\cite{han2020reduced}. As $K\to \infty$, the polygon approaches a disc and the corners approach type-IV vertices, for which the similarity solution has no distortion. This classification matches the numerical results on regular polygons as reported in Fig.~3 of~\cite{han2020reduced}.

We can use the similarity solutions above to understand the limiting solutions on isosceles triangles with apex angle $2\alpha$. The key difference between an isosceles triangle and an equilateral triangle is that the base vertices and the apex can be of different types (according to the classification in Table~\ref{tab:corner}) in an isosceles triangle. The similarity solutions constructed in~\eqref{eq:similarity} and the vertex classification in Table~\ref{tab:corner} only provide a local description of the nematic director $\mb n$, near each vertex. We can extrapolate these local solutions to the interior, to predict the qualitative properties of the unique solution of~\eqref{eq:lambda0} on arbitrary isosceles triangles, with apex angles $2\alpha$ and base angles $\frac{\pi}{2} - \alpha$, subject to tangent boundary conditions on the triangle edges. In contrast, the authors in~\cite{han2020reduced} analytically construct the solution of~\eqref{eq:lambda0} on regular polygons with Dirichlet boundary conditions, which is a global description on the entire polygon. The local observations on isosceles triangles can be summarized as follows: 
\begin{itemize}
    \item For acute isosceles triangles with $(0<2\alpha<90^\circ)$, all three vertices are type-I vertices with splay director profiles. The local splay profiles near the vertices, can be extended into the interior, predicting a trefoil director profile with an interior $-\frac{1}{2}$ defect. This is consistent with the limiting solution for an equilateral triangle, as constructed in~\cite{han2020reduced}.
    \item In the obtuse case with $(90^\circ<2\alpha<180^\circ)$, the apex is a type-III (bend) vertex whilst the base vertices are type-I (splay) vertices. We extend the local director profiles into the interior and predict a limiting bend-type or rotated director profile, that bends between the two base vertices.
    \item The transition between the trefoil and rotated solutions is mediated through the right isosceles triangle for which a defect line emanates from the type-II apex vertex.
\end{itemize}

\section{The large domain limit or the Oseen-Frank limit: the $\lambda\to\infty$ limit} \label{sec:largedomain}

We present some heuristics for the free energy~\eqref{freeenergy} in the regime $\lambda \gg 1$. As $\lambda \to \infty$, the minimizers of \eqref{freeenergy} tend to minimize the bulk potential almost everywhere, except near potential defects, etc. In other words, we expect the energy minimizers to respect the pointwise constraint
\begin{equation}
    4(q_1^2+q_2^2)=1, \quad\Rightarrow \quad s=+1   \label{soseen}
\end{equation}
almost everywhere on $\Omega$; this constraint also matches the Dirichlet boundary condition on $OMN$, except within the small neighbourhoods of the vertices. Following methods in~\cite{bethuel1994ginzburg,majumdar2010landau}, one can show that the minimizers of~\eqref{freeenergy} on the isosceles triangular domain $\Omega$, are of the form
\begin{equation}\label{Qmin}
\mb Q_{min} =  \left(\mb n_{min} \otimes \mb n_{min} - \tfrac{1}{2}\mb I \right)
\end{equation} 
where $\mb n_{min} = (\cos \varphi_{min}, \sin \varphi_{min})^T$ and $\varphi_{min}$ is a minimizer of
\begin{equation} \label{oseen}
I[\varphi] = \int_{\Omega} \frac{1}{2}|\nabla \varphi|^2 d^2 x
\end{equation} 
subject to Dirichlet conditions compatible with~\eqref{BC1}-\eqref{BC3}, almost everywhere. The minimizing director field, $n_{min}$, can have isolated interior point defects if the topological degree of the boundary datum is non-zero, and the defect set is typically a set of measure zero, denoted by $\mathcal{S}$~\cite{bethuel1994ginzburg, majumdar2010landau}. Hence, this can be viewed as a constrained minimization problem wherein the free energy \eqref{freeenergy} is minimized subject to the constraint $s=1$, everywhere away from $\mathcal{S}$. 
The order parameter $s$ vanishes inside $\mathcal{S}$.  Consequently, $\varphi_{min}$ is a harmonic function (almost everywhere), i.e., a classical solution of 
\begin{equation}
    \nabla^2 \varphi_{min} (x,y) = 0,  \quad (x,y)\in \Omega, \label{phieq}
\end{equation} 
subject to Dirichlet conditions compatible with~ \eqref{BC1}-\eqref{BC3}, away from $\mathcal{S}$. There are usually multiple choices for the Dirichlet conditions for $\varphi$, consistent with the imposed tangent boundary conditions, leading to multiple candidates for energy minimizers for $\lambda \gg 1$. 
\begin{figure}[h!]
\centering
\includegraphics[width=0.9\textwidth]{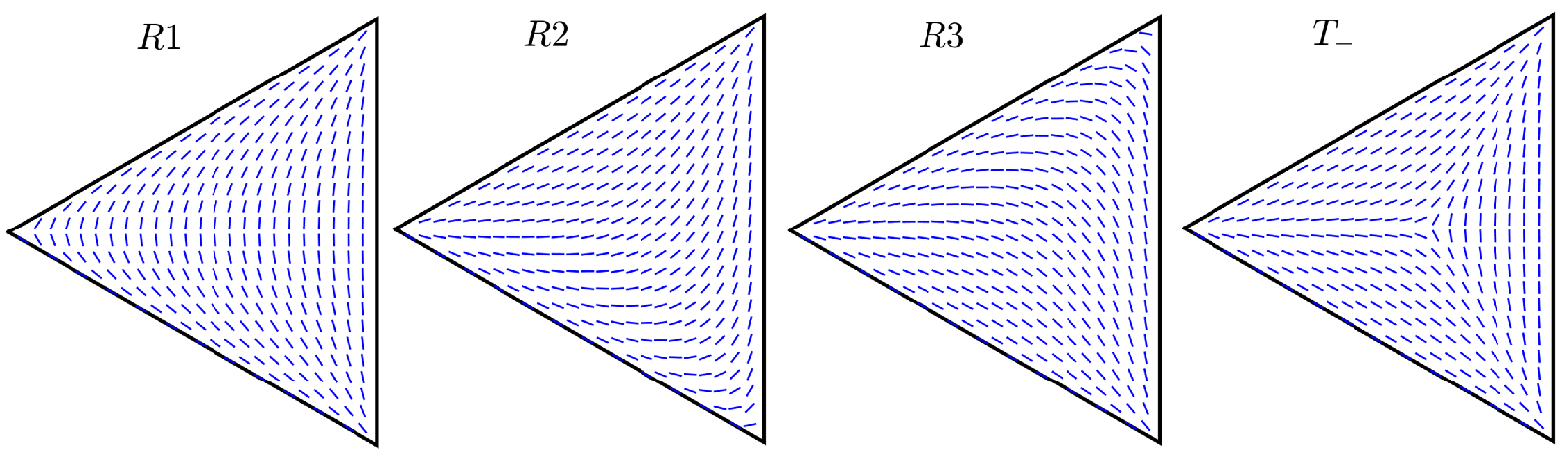} 
\includegraphics[width=0.37\textwidth]{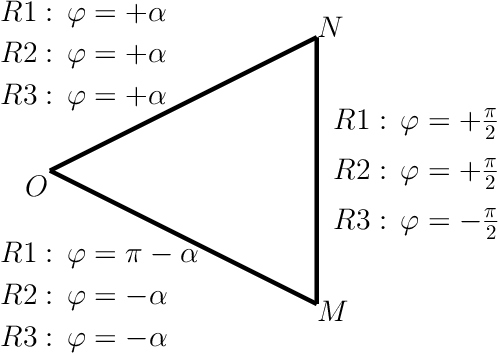} \hspace{1cm}
\includegraphics[width=0.37\textwidth]{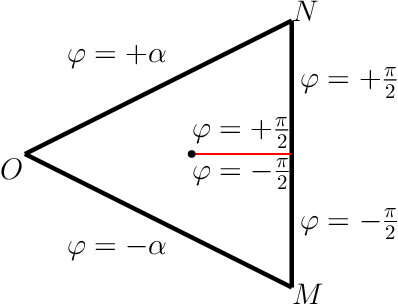} 
\caption{The three rotated solutions and the trefoil solution (top row) for an equilateral triangle, computed using~\eqref{phieq} complemented by the boundary conditions illustrated in the bottom row.} 
\label{fig:BCOF}
\end{figure}

In the remainder of this section, we focus on solutions of \eqref{phieq}, subject to Dirichlet conditions compatible with \eqref{BC1}-\eqref{BC3}, and use these solutions to qualitatively describe candidate profiles for $\mb Q_{min}$, as outlined above.
The tangent boundary conditions imply that $\mb n_{min}$ is tangent to each edge of $\Omega$, so that $\varphi_{min}= \alpha$ or $\varphi_{min}=\pi + \alpha$ on the edge $ON$, $\varphi_{min}=-\alpha$ or $\varphi_{min}=\pi - \alpha$ on the edge $OM$; and $\varphi_{min}=\pm \frac{\pi}{2}$ on the straight edge $MN$. 
By default, $\varphi$ is discontinuous at the vertices. Assuming that $\mathrm{deg}(\varphi_{min},\partial\Omega)=0$ and that there are no interior defects for $\varphi_{min}$, standard combinatorial arguments suggest that the corresponding director profile, $\mb n_{min}$, has two splay vertices and one bend vertex on triangular domains~\cite{han2020reduced}.  We label the three natural equilibria by \emph{rotated solutions}, denoted here by $R1$, $R2$, and $R3$ in Figure~\ref{fig:BCOF} (on an equilateral triangle). We compute these rotated solutions by solving \eqref{phieq}, subject to the boundary conditions outlined in the bottom left plot of Fig.~\ref{fig:BCOF}. The $R1$ solution has two splay vertices at the base vertices, whilst the apex vertex is a bend vertex. The rotated solutions $R2$ and $R3$ are mirror images of each other about $\theta=0$ and are energetically degenerate. The $R2$ and $R3$ solutions are distinguished by two splay vertices--one at the apex and one at the base--and the remaining base vertex is a bend vertex. In all three cases, the corresponding director profile, $\mb n_{min} = \left(\cos \varphi_{min}, \sin \varphi_{min}\right)$, bends between the two splay vertices. 

\begin{figure}[h!]
\centering
\includegraphics[width=0.47\textwidth]{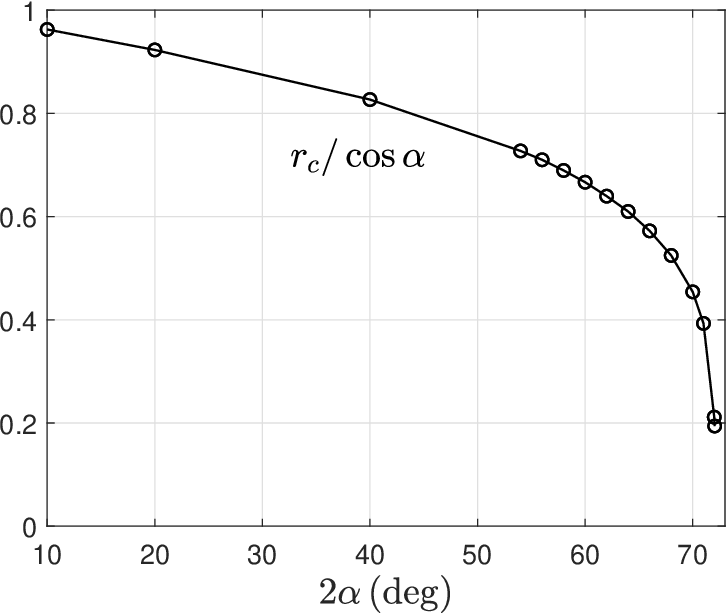} 
\includegraphics[width=0.47\textwidth]{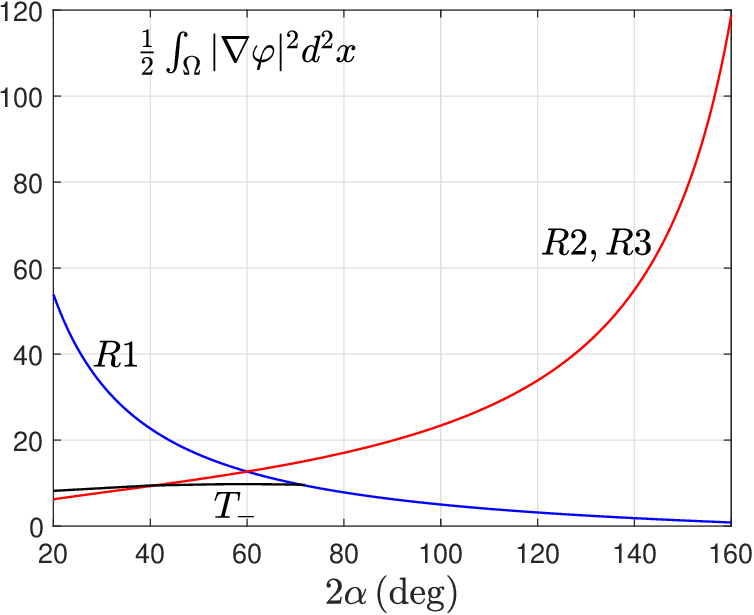} 
\caption{Left: Location of the interior $-\tfrac{1}{2}$ defect in the $T_-$ solution as a function of the apex angle $2\alpha$. Right: Numerically computed Oseen--Frank energy $\tfrac{1}{2}\int_\Omega|\nabla\varphi|^2d^2x$ for the four candidate minimizers as functions of $2\alpha$.} 
\label{fig:OFenergy}
\end{figure}

Besides the rotated solutions, the trefoil solution $T_-$ may be a candidate local minimizer of~\eqref{freeenergy} for $\lambda \gg 1$. We can construct a candidate trefoil solution with three splay vertices and an interior $-\tfrac{1}{2}$ defect, located at an interior point $P=(r_c,0)$ on the symmetry axis of the isosceles triangle. To numerically compute the $T_-$ solution using~\eqref{phieq}, a branch cut across which $\varphi_{min}$ jumps by $\pi$ needs to be introduced~\cite{miyazako2024defect}; the cut connects the interior $-\tfrac{1}{2}$ defect to a boundary point (see the bottom right plot of Figure~\ref{fig:BCOF}). We can calculate the optimal value of $r_c$ by the minimality condition,  $dI[\varphi_{min}(r_c)]/dr_c=0$, where $I[\varphi]$ is the Dirichlet energy defined above. For the equilateral triangle, symmetry implies that $r_c=1/\sqrt 3$, which places the defect at the centroid of the triangle.  Numerically computed values of $r_c/\cos\alpha$ as a function of the apex angle $2\alpha$  are plotted in the left plot of Figure~\ref{fig:OFenergy}. The numerical trends of $r_c$ indicate that the interior defect approaches the midpoint of the base $MN$ as $2\alpha\to 0$, while it approaches the apex $O$ at $2\alpha\approx 72^\circ$. Consequently, we speculate that the $T_-$ solution only exists for $2\alpha\leq 72^\circ$, for $\lambda \gg 1$. In contrast, in the $\lambda \to 0$ limit, the $T_-$ solution persists up to $2\alpha\leq 90^\circ$ and therefore, for finite $\lambda$, the corresponding existence threshold is expected to lie in the interval $[72^\circ,90^\circ]$.

In the right plot of Figure~\ref{fig:OFenergy}, we plot the Dirichlet energy (also known as the one-constant Oseen--Frank energy)~\eqref{oseen} of the four representative solutions, $R1$, $R2$, $R3$ and $T_-$, as functions of $2\alpha$; refer to Appendix A for the numerical evaluation of this integral since point defects (including the vertices) have infinite Dirichlet energy in two dimensions and we circumvent this energy blow-up by excising small neighbourhoods of point defects and vertices. From the energy trends, we observe that the $R1$ solution has the least energy for $2\alpha\geq 72^\circ$, whereas the two degenerate minima $R2$ and $R3$ are preferred for $2\alpha\leq 41^\circ$. This is consistent with the fact that $R1$ has a bend apex vertex and the bend profile is preferred for wide vertices. The $R2$ and $R3$ solutions have a splay apex vertex and splay director profiles are preferred for narrow vertices.  In the intermediate range, the $T_-$ solution has the minimum energy in this set of four competing solutions.

One can use the relations in~\eqref{Qmin} to construct candidate critical points/ minimizers of~\eqref{freeenergy} from $\varphi_{min}$, for $\lambda \gg 1$, as we have done here for the $R1$, $R2$, $R3$ and $T_-$ solutions. This construction can be generalised to give $\varphi_{min}$ profiles with multiple interior defects and arbitrary combinations of splay and bend vertices; we simply solve~\eqref{phieq} subject to appropriately defined Dirichlet conditions compatible with the prescribed vertex profiles and prescribed interior defects. However, it is unclear if these $\varphi_{min}$ profiles translate to critical points of~\eqref{freeenergy} (using the relation in~\eqref{Qmin}) and if so, whether the corresponding critical points are stable or not. In the following sections, we solve the full Euler--Lagrange equations associated with~\eqref{freeenergy} for finite but large values of $\lambda$, to check for the existence and stability of the $R1$, $R2$, $R3$ and $T_{-}$ solutions, as a function of $2\alpha$ and $\lambda$. % and hence, check the relevance of the asymptotic analysis in this section

\section{Illustrative numerical results for finite values of $\lambda$} \label{sec:numerical}

We now present illustrative numerical results for finite values of $\lambda$, by solving the full problem~\eqref{q1}-\eqref{BC3} along with the eigenvalue problem~\eqref{eigenvalue}. We focus on global minimizers of \eqref{freeenergy} as a function of $\lambda$ and $2 \alpha$, and indeed recover the $R1$, $R2$, $R3$ and $T_-$ solutions predicted by the similarity solutions and heuristic arguments, for $\lambda \ll 1$ and $\lambda \gg 1$ respectively. This does not preclude the existence of other local energy minimizers and unstable critical points of \eqref{freeenergy} on isosceles triangles, which will be investigated in sequential work. %, and future work will include detailed studies of solution landscapes on such canonical polygons to unlock their potential for confined nematic systems.

\begin{figure}[h!]
\centering
\includegraphics[width=0.9\textwidth]{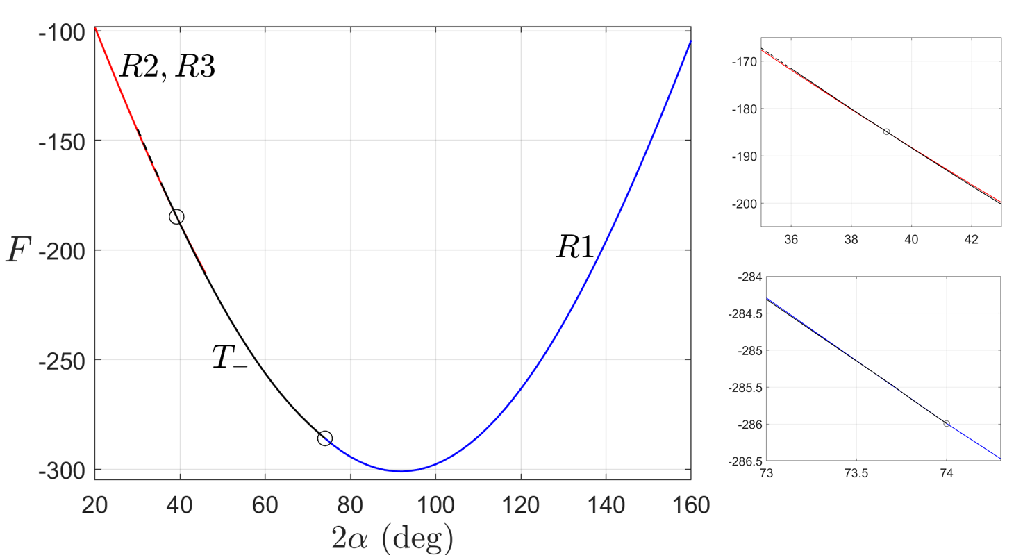} 
\caption{The free energy of the global energy minimizer in the set of $\left\{R1, R2, R3, T_- \right\}$ solutions for $\lambda=100$, and various apex angles. The circles locate points of crossover between different types of minimizers (see the two right insets for zoomed-in view). At the crossover, the $R1$ and $T_-$ solutions are indistinguishable and gradually transform from one to the other. The $T_-$ solution becomes unstable for $2\alpha<36^\circ$ indicated by dashed black line. The $T_-$ solution has the minimum energy for $39^\circ\leq 2\alpha \leq 74^\circ$, which is slightly wider than the range $41^\circ\leq 2\alpha \leq 72^\circ$ predicted by the large-domain limit analysis in the previous section (Figure~\ref{fig:OFenergy}).} 
\label{fig:energy}
\end{figure}

\begin{figure}[h!]
\centering
\includegraphics[width=0.9\textwidth]{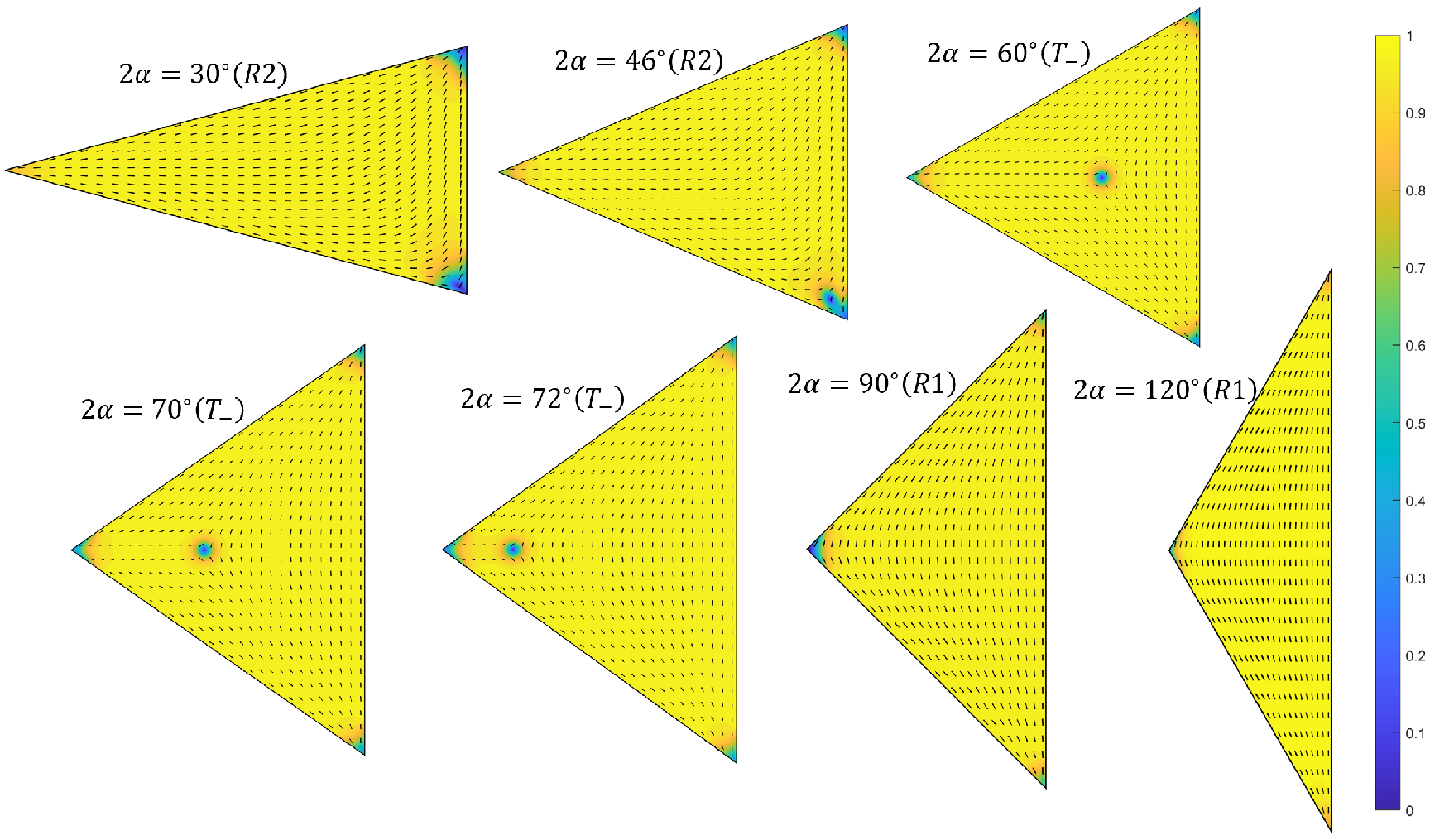} 
\caption{Representative numerical results for the nematic director profile and the scalar order parameter of critical points of \eqref{freeenergy}, subject to the tangent boundary conditions, for $\lambda=100$; also see Figure~\ref{fig:energy}.  These numerical solutions are stable except for the $2\alpha=46^\circ$ case, which pertains to an index-1 solution.} 
\label{fig:lam100plot}
\end{figure}

In Figure~\ref{fig:energy}, we compare the free energies \eqref{freeenergy} of the representative solutions $\left\{R1, R2, R3, T_{-} \right\}$ (when obtained numerically) and plot the energy of the global minimizer in this representative set, as a function of the apex angle $2\alpha$,  with $\lambda=100$. The global minimizer in this representative set is numerically stable; the stability is not guaranteed since we only compare energies within a restricted set of solutions and hence, needs to be checked. Note that $F<0$ because of the bulk energy term, which can be negative for large values of $s_+$. In Figure~\ref{fig:lam100plot}, we plot the director field and scalar order parameter for representative solutions of the full problem \eqref{q1}-\eqref{BC3}, for different values of $2\alpha$ as in Figure~\ref{fig:energy}. For $2\alpha \lessapprox 39^\circ$, the ground state is degenerate and corresponds to either the $R2$ or $R3$ solution, for which the apex vertex is a splay vertex and the director field (modelled by $\mb n$ in \eqref{eq:Q}) bends around one of the vertices. For $2\alpha \gtrapprox 74^\circ$, the global energy minimizer is the $R1$ solution, for which the apex vertex is a bend vertex and the director field smoothly interpolates between the two base splay vertices. This is consistent with the predictions  in Table~\ref{tab:corner}, for vertex profiles according to their vertex angles. In the intermediate range, the $T_-$ solution is the global energy minimizer, with three splay vertices and an interior $-\frac{1}{2}$ defect.

The crossover between the $T_-$ and $R1$ solutions is gradual. This transformation is mediated by the migration of the interior $-\tfrac{1}{2}$ defect in the $T_-$ solution, towards the vertex $O$ as $2\alpha$ increases (recall that such a transformation is predicted at $2\alpha=90^\circ$ for $\lambda=0$). Near the crossover point, the apex vertex $O$ is of a splay character and is accompanied by a slightly displaced $-\tfrac{1}{2}$ defect in the interior, as seen for $2\alpha = 72^\circ$. As $\alpha$ increases, the splay vertex and the $-\tfrac{1}{2}$ defect coalesce to yield a bend profile at the apex, and the $T_-$ solutions give way to the $R1$ solution in Figure~\ref{fig:energy}. On the other hand, the transition between the $R2/R3$ and $T_-$ is not continuous. As indicated in Figure~\ref{fig:energy}, the global minimizer switches from the $R2/R3$ (with a unique bend vertex) solutions to the $T_-$ solution at approximately $2\alpha=39^\circ$. The $R2/R3$ configurations persist as local minimizers in the interval $39^\circ\leq 2\alpha\leq 45^\circ$, but become unstable at $2\alpha=46^\circ$, beyond which the rotated solution could not be numerically continued. At $2\alpha=46^\circ$, while the trefoil solution exhibits an interior $-\tfrac{1}{2}$ defect located on the symmetry axis, the rotated solution develops a composite structure near a base vertex as shown in Figure~\ref{fig:lam100plot}. Unlike the smooth defect migration along the symmetry axis observed near $2\alpha\approx 72^\circ$, the instability here may be attributed to a strongly asymmetric rearrangement of the $-\tfrac{1}{2}$ defect as it attempts to detach from the base bend vertex and move into the interior with increasing $2\alpha$.

%A similar composite structure appears near vertex $M$ for $2\alpha = 46^\circ$, when the rotated solution gives way to the trefoil solution. Here, there is a unique bend vertex for the rotated solution. The bending is expelled into the interior, creating a splay vertex and an interior $-\frac{1}{2}$ defect near the transformed vertex. This relieves the geometric frustration as $2\alpha \to 46^\circ$. As $2\alpha$ increases, the created or expelled $-\frac{1}{2}$ defect migrates to the triangle interior, yielding three splay vertices and the $T_{-}$ solution. %In the limiting case, the $-\frac{1}{2}$ defect migrates to the bend vertex, yielding a splay vertex and the $T_{-}$ solution appears (with three splay vertices).  

\begin{figure}[h!]
\centering
\includegraphics[width=0.49\textwidth]{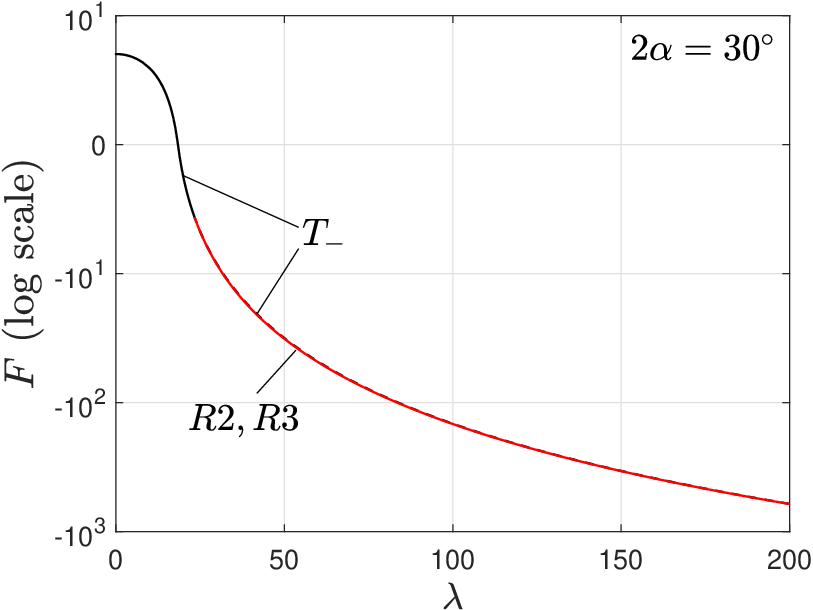}
\includegraphics[width=0.49\textwidth]{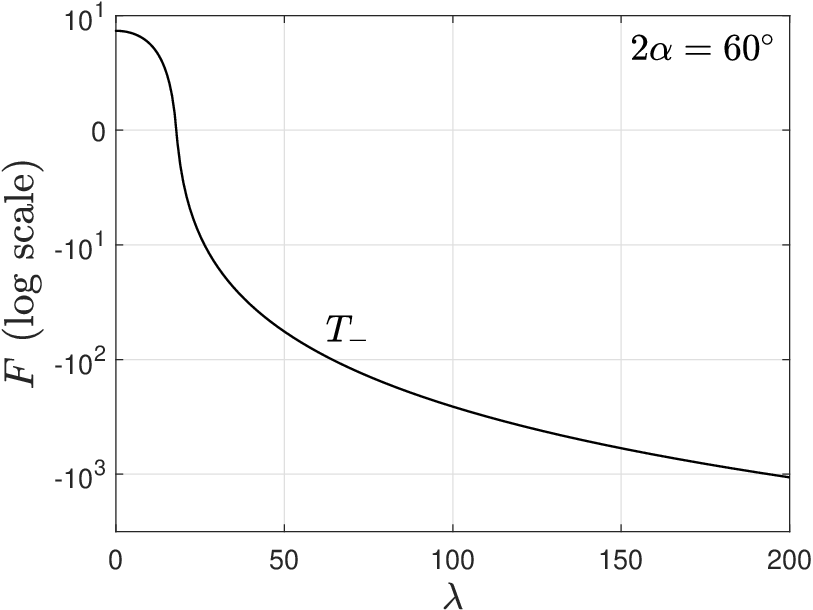}
\includegraphics[width=0.49\textwidth]{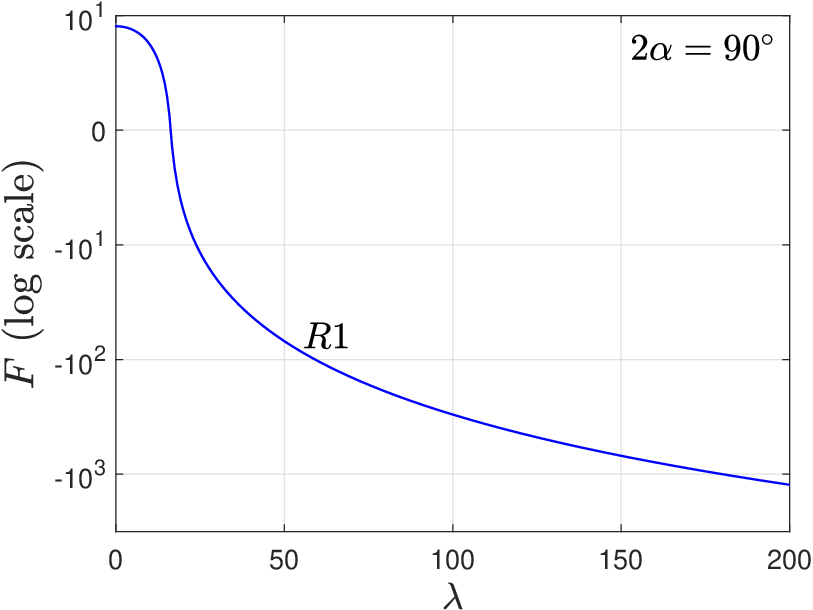}
\caption{Energy trends as a function of $\lambda$ for three apex angles corresponding to the $30^\circ$-$75^\circ$-$75^\circ$ triangle, an equilateral triangle and the right isosceles triangle. In the first case, the $T_-$ solution becomes unstable for $\lambda>23$ (denoted by dashed black line which lies a little above the solid red line) through a supercritical pitchfork bifurcation.} 
\label{fig:bifur}
\end{figure}

Such a composite structure has also been noted  in~\cite{yao2022defect}, who refer to the slightly displaced $-\tfrac{1}{2}$ defect as a \textit{hidden defect}. This composite structure points to a critical corner angle $\theta_c$, such that the director field of global energy minimizers is simply unable to bend if the corner angle is less than $\theta_c$. In this case, the director field relieves itself of geometric frustration by the presence of a nearby additional $-\tfrac{1}{2}$ defect. We explain this heuristically by comparing the topological charges of splay and bend corners, measured in terms of the rotation of the director field near the vertex. Since a splay corner rotates the director angle by $-2\alpha'$ and a bend corner rotates it by $\pi-2\alpha'$, the corresponding charges are given by
\begin{equation}
    +1 \quad \text{for splay corners  and} \quad 1-\frac{\pi}{2\alpha'} \quad \text{for bend corners}.
\end{equation}
The splay corner charge remains constant while the bend corner charge grows rapidly as the corner becomes sharper, i.e., as $2\alpha' \to 0$. This makes bend corners energetically unfavourable for sufficiently small angles. In particular, the magnitude of the bend-corner charge is greater than that of the splay corner whenever $2\alpha'<90^\circ$. However, a simple criterion such as $2\alpha'<90^\circ$ is not sufficient to predict whether a bend corner will be avoided, because one must also account for the energetic cost of an interior $-\tfrac{1}{2}$ defect and the frustration induced by the boundary conditions. In other words, there is a complex trade-off between the bend-corner energy, the interior defect energy and the boundary conditions that determines the apex angle for which director bending becomes favourable. For $\lambda=100$, this balance is attained near the apex vertex $O$ when the corner angle is approximately $74^\circ$, and near the base vertices $M$ or $N$ when the base angle is about $90^\circ-39^\circ/2=70.5^\circ$. The slight difference between these critical angles may be attributed to symmetry: the bending pattern at $O$ is symmetric about the corner bisector whereas this symmetry is absent at the other two base vertices.  Informally, we speculate that the global energy minimizer of \eqref{freeenergy} prefers a splay profile near a corner, if the corner angle is less than about $70^\circ$ for $\lambda=100$. In such cases, interior defects may appear to relieve the energetic penalty associated with sharp bend corners. For example, the $T_-$ solution is the unique global energy minimizer on an equilateral triangle for all $\lambda$, because all corners are sharp enough to favour a splay profile and an interior defect, to avoid the energetic cost of director bending around corners.  

To complement the preceding results for the fixed $\lambda=100$, we present the energetics of the global minimizers as a function of the domain size $\lambda$ in Figure~\ref{fig:bifur}, for three representative apex angles. For the equilateral triangle and the right isosceles triangle, the global minimizers correspond to the $T_-$ and $R1$ solutions, respectively, for all values of $\lambda$. For the $30^\circ$-$75^\circ$-$75^\circ$ triangle, the global minimizer is the $T_-$ solution when $\lambda<23$ and switches to the $R2$ or $R3$ solution for larger values of $\lambda$. As $\lambda$ increases, the interior fractional defect in the $T_{-}$ solution migrates towards and coalesces with one of the base vertices, so that the base vertex changes from splay to bend type, yielding either the $R2$ or $R3$ solution. The $T_-$ solution loses stability at $\lambda=23$ through a supercritical pitchfork bifurcation, giving rise to the stable rotated solutions for $\lambda>23$. On these grounds, we speculate that there is greater scope for multistability and richer solution landscapes (with multiple critical points of \eqref{freeenergy}) on isosceles triangles with acute apex angles, compared to isosceles triangles with $2\alpha \geq 90^\circ$; this warrants further study in the future.

\section{Concluding remarks}\label{sec:conclusions}

In this paper, we have modelled NLC equilibria on isosceles triangles, subject to tangent boundary conditions, in the continuum rLdG framework. Triangles are the simplest polygons and it is essential to understand how symmetry, geometric frustration, size and anchoring couple to each other and determine properties of NLC equilibria. Equilateral triangles have been studied to some extent in \cite{han2020reduced} and we include the effects of symmetry breaking, by extending the work in \cite{han2020reduced} to isosceles triangles. We have two notable findings in this context, both of which are intuitive but can be of value in comprehensive studies of 2D and 3D NLC equilibria.

We study rLdG equilibria, modelled by minimizers of the rLdG free energy, in two asymptotic limits: the $\lambda \to 0$ limit valid for very small domains and the $\lambda \to \infty$ limit, valid for very large domains. Previous work indicates that the asymptotic results in the $\lambda \to 0$ limit often survive for $\lambda\leq 10$ or nano-scale domains. In the $\lambda \to 0$ limit, the rLdG Euler--Lagrange equations reduce to the Laplace equations for the rLdG $\Qvec$-tensor order parameter. We do not solve the limiting equations explicitly. Instead, we classify each vertex as being \emph{splay}, or \emph{bend} or \emph{singular} according to the vertex opening angle (see Table~\ref{tab:corner}). The local vertex profiles can be extrapolated to predict the qualitative properties of the limiting solutions or the unique NLC equilibrium profile in the $\lambda \to 0$ limit. For example, we predict that if all vertex angles are less than $90^\circ$, then the limiting solution has three splay vertices accompanied by an interior $-\tfrac{1}{2}$ defect, referred to as the \emph{trefoil} solution. The location of the interior fractional defect depends on $2\alpha$; indeed, numerical results suggest that the interior fractional defect migrates from the base to the apex, as the apex angle approaches a right angle. 

For non-zero values of $\lambda$, the rLdG Euler-Lagrange equations are a nonlinear and coupled system of two elliptic partial differential equations, subject to Dirichlet conditions in our rLdG setting. In the $\lambda \to \infty$ limit (relevant for large micron-scale geometries), we argue that the minimizers of the rLdG free energy can be approximated uniformly by $\Qvec_{min}$ in \eqref{Qmin} (almost everywhere except for a set of measure zero), where $\Qvec_{min}$ is a minimizer of the bulk potential and the corresponding director, $\mb n_{min}$, is fully determined by a Dirichlet boundary value problem on the isosceles triangle. There are multiple choices of $\mb n_{min}$ compatible with the tangent boundary condition, opening avenues for multistability.

We identify four candidates for rLdG energy minimizers in the $\lambda \to \infty$ limit on grounds of minimal topological charge and symmetry: three \emph{rotated} solutions with two splay vertices and one bend vertex, and the \emph{trefoil} solution with an interior fractional defect. This is not a proof, but simply a construction based on asymptotic arguments and heuristics. Unlike an equilateral triangle, the rotated solutions are not energetically degenerate on isosceles triangles. We numerically compute the rotated solutions and the trefoil solution for large values of $\lambda$, by solving the full nonlinear rLdG Euler--Lagrange equations subject to the Dirichlet tangent boundary conditions. The numerically computed solutions are stable. We do not find competing stable solutions or multiple stable solutions (global or local) for most choices of $\lambda$ and $2\alpha$. For example, we find that the rLdG energy minimizer in the restricted set $\left\{R1, R2, R3, T_-\right\}$ transitions from the trefoil solution to one of the rotated solutions, as $\lambda$ increases, for small $2 \alpha$. However, we do not numerically observe coexisting stable rotated and trefoil solutions, except within a narrow range of apex angles, $36^\circ\leq 2\alpha< 45^\circ$ when $\lambda=100$. Similarly, for fixed $\alpha$, we expect to find co-stability of $R2/R3$ solutions and the $T_{-}$ solution over a small range of $\lambda\in \left[\lambda_1, \lambda_2 \right]$, where $\lambda_1$ is sufficiently large. We only find the stable trefoil solution on the equilateral triangle, for all $\lambda$ under consideration, whilst the energetically degenerate rotated solutions are reasonable candidates for local energy minimizers (for large $\lambda$). This could be an artefact of the Dirichlet boundary conditions. In \cite{rajamanickam2026strong} and \cite{luo2012multistability}, the authors formulate a more physically grounded tangent boundary condition near the vertices, as opposed to a linear interpolation employed in our paper (see \eqref{BC1}-\eqref{BC2}). These physically grounded boundary conditions may be better suited to capture multistability. One could also use surface energies on the edges such that the surface anchoring coefficient decays at the vertices, since it is likely very difficult to enforce strong anchoring at sharp corners.

Importantly, we can construct a zoo of candidates for $\Qvec_{min}$ as in \eqref{Qmin} by solving Dirichlet boundary value problems for $\phi_{min}$ in \eqref{phieq}. There are numerous admissible choices of the boundary conditions, compatible with the tangent boundary conditions and any number of prescribed interior defects. There are at least $8$ admissible choices of $\phi_{min}$ on the triangle edges themselves. The question then is - are there universal rules that determine which of the artificially constructed $\Qvec_{min}$ survive as critical points of the rLdG energy and which survive as stable critical points of the rLdG energy, for large but finite values of $\lambda$? How does the Morse index of an unstable saddle point of the rLdG free energy depend on the choice of $\phi_{min}$, or on the location and multiplicity of interior defects? Some answers may lie in the concept of the renormalised energy in the Ginzburg--Landau theory for superconductivity but this question may hold clues to designing self-assembly and multistability at will, for the next generation of NLC-type soft materials.

\bmhead{Acknowledgements}
 P.R. acknowledges support from Leverhulme Research Project Grant RPG-2021-401. Part of this work was carried out while P.R. and A.M. were at the University of Strathclyde.
 A.M. acknowledges support from a Visiting Professorship at the University of Strathclyde and a Leverhulme Research Project Grant RPG-2021-401. On behalf of all authors, the corresponding author states that there is no conflict of interest.

\bibliography{sn-bibliography}% common bib file
%% if required, the content of .bbl file can be included here once bbl is generated
%%\input sn-article.bbl

\section*{Appendix A: Numerical methods}

All numerical computations in this study are carried out using COMSOL Multiphysics, which employs the finite element method for spatial discretization and the MUMPS (Multifrontal Massively Parallel Sparse) direct solver. The mesh size used in the simulations lies in the range $(1/300,1/250)$. In particular, the various critical points are computed by solving the problem~\eqref{q1}-\eqref{BC3} with different initial guesses. Throughout the numerical simulations, the corner regularisation parameter $\epsilon$ is fixed at $0.1$.

The numerical evaluation of the Oseen--Frank energy defined in~\eqref{oseen} requires additional care. Within the Oseen–Frank framework, the gradient $\nabla\varphi$ diverges at defect locations, leading to singular contributions to the energy. To handle this issue, the singular part is excluded by removing a small disc of radius $0.01$ centred at each defect point. This exclusion radius is chosen to be larger than the mesh size, which is approximately $1/300$, ensuring numerical consistency.

\end{document}